\documentclass[conference]{IEEEtran}
\IEEEoverridecommandlockouts
\usepackage{cite}
\usepackage{amsmath,amssymb,amsfonts}
\usepackage{algorithmic}
\usepackage{graphicx}
\usepackage{textcomp}
\usepackage{xcolor}
\usepackage{subcaption}
\usepackage[ruled]{algorithm2e}
\usepackage{hyperref}

\def\BibTeX{{\rm B\kern-.05em{\sc i\kern-.025em b}\kern-.08em
    T\kern-.1667em\lower.7ex\hbox{E}\kern-.125emX}}
\begin{document}

\title{Trustworthy misinformation mitigation with soft information nudging}

\author{\IEEEauthorblockN{Benjamin D. Horne}
\IEEEauthorblockA{\textit{Department of Computer Science} \\
\textit{Rensselaer Polytechnic Institute}\\
Troy, NY, USA\\
horneb@rpi.edu}
\and
\IEEEauthorblockN{Maur\'{i}cio Gruppi}
\IEEEauthorblockA{\textit{Department of Computer Science} \\
\textit{Rensselaer Polytechnic Institute}\\
Troy, NY, USA\\
gouvem@rpi.edu}
\and
\IEEEauthorblockN{Sibel Adal{\i}}
\IEEEauthorblockA{\textit{Department of Computer Science} \\
\textit{Rensselaer Polytechnic Institute}\\
Troy, NY, USA\\
adalis@rpi.edu}
}

\maketitle

\begin{abstract}
Research in combating misinformation reports many negative results:
facts may not change minds, especially if they come from sources that
are not trusted. Individuals can disregard and justify lies told by
trusted sources. This problem is made even worse by social recommendation algorithms which help amplify conspiracy theories and
information confirming one's own biases due to companies' efforts to
optimize for clicks and watch time over individuals' own values and
public good. As a result, more nuanced voices and facts are drowned
out by a continuous erosion of trust in better information sources. Most misinformation mitigation techniques assume that discrediting, filtering, or demoting low veracity information will help news consumers make better information decisions. However, these negative results indicate that some news consumers, particularly extreme or conspiracy news consumers will not be helped. 

We argue that, given this background, technology solutions to
combating misinformation should not simply seek facts or discredit
bad news sources, but instead use more subtle nudges towards better information consumption.  Repeated exposure to such nudges can help promote trust in better information
sources and also improve societal outcomes in the long run. In this
article, we will talk about technological solutions that can help us
in developing such an approach, and introduce one such model called Trust Nudging.

\end{abstract}


\section{Introduction}
There are many useful and necessary paths to combating misinformation. These paths include technical methods to identify incorrect or misleading claims~\cite{popat2016credibility, hassan2017claimbuster}, methods to make correct information more easily available~\cite{barron2019proppy}, and methods to identify sources that disseminate incorrect information~\cite{horne2018accessing, baly2018}. Some research pathways are non-technical, but equally if not more important, as they address the underlying issues and institutions that lead to the creation, dissemination, and consumption of misinformation~\cite{starbird2018ecosystem, starbird2019disinformation1}. There has also been significant growth in political fact-checking organizations, including the fact-checking of news articles, social media posts, and claims made by politicians~\cite{amazeen2015sometimes}. 

In all these paths, there are various underlying challenges. Overall, the misinformation problem is deeper than identifying what is ``fake
news.'' While the dissemination of proven incorrect information is a real problem, confirming that specific information is incorrect can be deeply contextual. This opens up the information's correctness to debate which can lead to suppression of minority voices and opinions. Furthermore, sophisticated misinformation campaigns mix correct and incorrect information, which can cause uncertainty and make discrediting information more difficult. The majority of technical solutions have focused on classifying these extremes (fake and real), which leaves automatic assessment of uncertain, mixed veracity, and deeply contextual information difficult to assess. 

A deeper problem that is left unaddressed in the technical research threads is what to do when information corrections, whether done by an algorithm or journalist, do not work. Even if information is correctly discredited, consumers may choose to ignore the correct information, due to distrust in the platform, algorithm, or organization providing the corrected information. This behavior is particularly prevalent among consumers with extreme or conspiratorial views~\cite{sunstein2009conspiracy}. If low veracity information is filtered out or demoted, consumers may become more extreme and distrust the contemporary media platforms. The rise of alternative ``free speech'' platforms such as Gab and Bitchute are examples of this~\cite{zannettou2018gab}. Similarly, if consumers perceive this filtering, demoting, or discrediting as partisan, distrust in information corrections can persist~\cite{turcotte2015news, hanitzsch2018caught}, resulting in reduced trust for the news source, the platform/algorithm curating information and the fact-checking organization. Due to this distrust, solutions to correcting misinformation can be ineffective for some consumers~\cite{sunstein2009conspiracy}. 
  
In this paper, we begin to address this problem: {\em How can online media systems, if they are willing to, support trust in higher quality sources?} Specifically, we propose \texttt{Trust Nudging}, a generic, trust-based recommendation model for improving the quality of news consumed. This proposed method is built on the concept of nudging, which provides {\em alternatives without forbidding any options or significantly changing the economic incentives}~\cite{thaler2009nudge}. In essence, we would like to provide alternative sources of information to users at decision points without taking away any agency from them and without suppressing information. To do this, we provide subtle recommendations to readers in order to nudge them towards news producers of objectively higher quality, but also have a chance of being trusted; thereby avoiding recommendations that may not work or may break trust. We leverage news relationship graphs and the news already read by the consumer to approximate the trust of recommended news sources. Using a simulation, we show that this model can slowly increase the quality of news a consumer is reading, while not demanding a substantial shift in trust or ideological beliefs. Furthermore, we show that, as a side effect, this model lessens the partisan extremity of the news being read. In addition to simulating this generic model, we outline different research threads that can help support this approach, as well as, different choice architectures that can support better information consumption. Lastly, we discuss the benefits and potential drawbacks of this type of recommendation method.

\section{Related Work}
\subsection{Current Approaches to Misinformation Mitigation}~\label{sec:current}
There have been many proposed technical solutions to combating online misinformation. The vast majority of the technical solutions have been developed as detection systems, which can filter out or discredit news that is of low veracity or provide fact-checking on the claims within the media. These solutions range widely in terms of technical methods used, including various types of machine learning models using the content in news articles and claims~\cite{horne2018accessing, popat2016credibility, guacho2018semi}, deep neural network models utilizing social features of shared news~\cite{ruchansky2017csi}, source-level ranking models~\cite{barron2019proppy}, and knowledge-graph models for fact-checking~\cite{hassan2017claimbuster}. Many of these approaches have shown high accuracy in lab settings. Some approaches have also shown robustness to concept drift and some adversarial attacks~\cite{horne2019robust}. 

The assumption of detection-based misinformation solutions is that discrediting false or misleading information will help consumers make fully informed decisions. However, there is reason to believe that discrediting or filtering out bad information will not help some news consumers. First, discrediting information may be ignored by consumers with extreme or conspiratorial views. As described in~\cite{sunstein2009conspiracy}: ``A distinctive feature of conspiracy theories is their self-sealing quality. Conspiracy theorists are not likely to be persuaded by an attempt to dispel their theories; they may even characterize that very attempt as further proof of the conspiracy.'' This negative reaction is related to the ``back-fire effect,'' where the consumer's beliefs become even stronger when information about strongly-held beliefs is discredited~\cite{nyhan2010corrections}, although this effect does not consistently occur~\cite{wood2016elusive}. Additionally, it has been shown that discrediting misinformation does not always help reasoning, making the usefulness of discrediting information even more nuanced. Specifically, future reasoning can be influenced by misinformation even if that information has been corrected or debunked (often called the continued-influence effect)~\cite{johnson1994sources}. This lingering effect can appear as recalling the incorrect information from memory, even if it was corrected immediately~\cite{ecker2017reminders}, or maintaining strongly-held attitudes about the topic, even if beliefs are correctly updated~\cite{swire2019they, thorson2016belief}. These effects not only exist for explicit misinformation, but also more subtle, implied misinformation in news articles~\cite{rich2016continued}. The corrective effects of fact-checking can also be limited if they are given with unrelated contextual information~\cite{garrett2013undermining}.

Second, information trust plays a role in belief updating.
Information trust is a mix of an individual's own judgment of information and the trust in the source~\cite{Adali:2013, Hilligoss:2008}. When assessing trust, an individual may rely solely on their own evaluation of the information, especially if the source is not trusted, in which case confirmation of the reader's own beliefs as well as other heuristics may play a large role~\cite{lewandowsky2012misinformation}. For example, information that is compatible with a person's current beliefs can be seen as more credible and stories that are coherent may be easier to trust~\cite{lewandowsky2012misinformation}. Many sources disseminating misinformation have become quite good at targeting such heuristics~\cite{horneadali2017}. Similarly, for trusted sources, the information can be accepted as true without much critical evaluation. Trust for sources is a complex concept as well, evaluated on multiple axes, such as the alignment of the source's values with the reader or their perceived competence.  

Over the past decade there has been an erosion of trust in media and political institutions~\cite{turcotte2015news, hanitzsch2018caught}, which can materialize as the polarization of trust in news outlets. If an algorithm recommends news from a high quality source that is initially distrusted by the consumer, it is unlikely the consumer makes a change. In the context of politics, a strongly partisan reader may only trust sources closely aligned with their political view. In this case, recommending an article from the opposite political camp is highly unlikely to work. Similarly, telling the reader of a conspiratorial news source to read an article from a neutral source is unlikely to yield any impact. As both disinformation production and information trust become more politically polarized, methods that filter, block, demote, or discredit may be less effective, as they may be perceived as partisan paternalism. 

The decline of trust in long-standing news outlets is matched with an increase in trust of information recommended on social media~\cite{turcotte2015news}, although as social media platforms become more contemporary, this trust has also wavered. Partisan-based trust in information from social media is concerning as disinformation is often partisan and more prevalent on social media~\cite{faris2017partisanship}. Furthermore, a great deal of research and discussion shows that social media recommendation systems further complicate this problem~\cite{ribeiro2019auditing}. For example, Facebook's news feed algorithm has been said to ``rewarded publishers for sensationalism, not accuracy or depth\footnote{\url{www.wired.com/story/inside-facebook-mark-zuckerberg-2-years-of-hell/}}.'' As a result, news sources focused on providing complete, neutral, and nuanced commentary on factual events may end up being demoted in news feeds, providing passive consumers with little opportunity to develop trust for these high quality sources.

Third, filtering out, blocking, or demoting bad information can be perceived as loss of agency or suppression of free speech, which may increase polarization, particularly for those consumers with conspiratorial views. One very prominent example of this is the rise of alternative media platforms such as Gab, Bitchute, and Voat~\cite{zannettou2018gab}, which harbor conspiracy theorist and hyper-partisan information producers. These platforms self-proclaim that they have been created to promote free speech rights that have been taken away from them through the demonetization and removal from contemporary platforms like Twitter and YouTube. While the movement of partisan and conspiracy media from mainstream platforms to alternative platforms may stagnate misinformation flow to the wider-public (which is still up for debate), it also creates even more extreme echo chambers, which can lead to radicalization~\cite{ribeiro2019auditing}.

These negative results do not necessarily invalidate the applicability of current approaches. The active engagement of social scientists are needed to understand how and when employing such methods are necessary to reduce potential harm to individuals especially on social media. There is evidence that fact-checking efforts discourage politicians from lying and can, under the right conditions, change consumer beliefs~\cite{amazeen2015sometimes}. There is also evidence that automated methods help consumers determine if a news article is unreliable or biased, particularly with feature-based explanations~\cite{horne2019rating}. However, the benefits are not uniform: users who read or share news on social media benefit less from these type of explanations. Lastly, there is evidence that flagging false news can reduce false news sharing~\cite{mena2019cleaning}, which suggests that filtering out or demoting maliciously false content can prevent its spread to the wider-public. Despite these positive results, there are also ample negative results which should be addressed. This paper proposes one idea to begin addressing them. 

\subsection{Nudges and Choice Architectures}
Highly related to the method proposed in this paper is the concept of nudging. The concept of nudging has been well studied in the field of Behavioral Economics. The idea was particularly popularized by Richard Thaler and Cass Sunstein's 2008 book entitled \textit{Nudge: Improving Decisions About Health, Wealth, and Happiness}~\cite{thaler2009nudge}. According to Thaler and Sunstein, a nudge is:\\

\begin{quote}
``any aspect of the choice architecture that alters people's behavior in a predictable way without forbidding any options or significantly changing their economic incentives. To count as a mere nudge, the intervention must be easy and cheap to avoid. Nudges are not mandates. Putting fruit at eye level counts as a nudge. Banning junk food does not.'' \\ 
\end{quote}

In other words, nudges are small changes in the choice architecture that work with cognitive biases in decision making. The choice architecture is simply the environment in which a decision is being made, and that decision is nudged by this environment, even if it is not purposely designed to do so. Being nudged by a choice architecture is unavoidable. A classic example is the ordering of items in a list, where the items atop the list are often selected more than any other item in the list, even if the ordering is arbitrary~\cite{salganik2006experimental}. Another example is keeping the status quo. If something is set by default (i.e. the ringtone on a phone) it is unlikely to be changed~\cite{thaler2009nudge}. Because these subtle, but influential, environmental nudges are unavoidable, Thaler and Sunstein argue that these environments should be purposely designed to make nudges beneficial~\cite{thaler2009nudge}.

There are situations where well-designed nudges may be particularly helpful. People tend to make bad choices when a situation's cost are not realized until later, the decision is difficult to assess, the choice being made is infrequent, or there is no immediate feedback~\cite{thaler2009nudge}. In these cases, a well-designed nudge can benefit the decision maker. 

Consuming online news cuts across many of these situations. Often consuming information has no immediate impact on the reader or sometimes the direct impact on the individual consumer is never realized, as its impacts may be on a societal level. In some cases, there is a direct impact on the individual consumer, but it is delayed. For example, a news consumer may choose not to vaccinate their child due to false information on vaccination safety. After some time, their child may catch a preventable disease due to this misinformed decision. Because of this delayed and sometimes indirect feedback from information consumption, it may be difficult to translate the costs and benefits of news reading decisions to the consumer. Additionally, as previously discussed in Section~\ref{sec:current}, humans use many mental shortcuts when assessing information. The choice architecture of online news consumption often exacerbates the use of these shortcuts. One such example is the passive nature of scrolling through a social media feed. Rather than actively seeking information about current events, users are passively consuming this information, sometimes piecemeal, creating potentially incorrect factoids when recalling reported events. This infinite-scrolling design choice nudges users to consume passively. 

There are many ways current news consumption systems, like social media, could be improved through better choice environments and nudges. In fact, many of these design changes are quite simple, but may oppose the current profit-driven architectures of social media platforms. Examples include: 
\begin{itemize}
    \item Limiting infinite scrolling or auto-play features, to nudge users to more actively consume information,
    \item Setting the default news feed for new user accounts to contain reliable and diverse information sources, utilizing status quo bias~\cite{thaler2009nudge},
    \item Show diverse alternatives next to a given news article in a news feed, providing an opportunity for the user to choose higher quality or diverse information,
    \item Provide a portion of the body text with the title of an article in news feed, avoiding ``implied misinformation.~\cite{rich2016continued}'.
\end{itemize}

In many of these systems, there is an additional layer in the form of recommendation or sorting, which explicitly influences what news is consumed (often much more explicitly than one would consider a ``nudge'' to be). In a sense, these recommendation systems are filtering down the choices of information to consume, in the same way Netflix filters down our movie choices based on what we previously watched. However, while recommendation about what movie to watch is rather benign, recommending what information to consume may not be. Information that is highly engaged with is not necessarily information of high quality~\cite{vosoughi2018spread}. Hence, removing the engagement-based recommendation systems for news consumption alone may be enough to nudge better consumption. 

\section{Using Trustful Nudges in News Recommendations}~\label{sec:usingtrustnudge}
Given the potential downfalls of discrediting partisan information and the complexity of information trust, we propose using ``nudges'' to help consumers make better consumption and sharing decisions. Specifically, we propose a trust-based recommendation model for news quality called \texttt{Trust Nudging}. The model's goal is to provide subtle recommendations to readers in order to nudge them towards news producers of objectively higher quality, without demanding a substantial challenge to one's beliefs. At a high level, suppose we are given two partial orderings of information sources, one along a trust/belief axis and one along a nonpartisan quality axis. The objective is given an article $A$, find an alternative article $B$ that is higher in quality than $A$ and has some proximity to $A$ along the trust dimension. The trust nudge is the act of providing to the user article $A$ and $B$ together in any point where they are making a decision, such as to share or to read. Users have the choice to read or share one, both or none of the articles. Both the quality dimension and the trust dimension can be approximated in many ways and be computed at different granularity. We present one approach in this paper. 

To implement the Trust Nudging model, we need several pieces of information:
\begin{enumerate}
    \item A relationship graph between news sources, which is used to approximate the likelihood that a user will trust a given news source.
    \item An approximate ground truth of both quality and political leaning for each news source in the relationship graph. This is used in both in the trust calculation and to ensure recommendations are of higher quality than the users' current consumption. 
    \item A set of user reading profiles, which simply state what news sources a user reads or trusts. In real life this could be the news a user is exposed to in his social media feed. 
\end{enumerate}

The key to implementing the model is the news source relationship graph. While theoretically this relationship graph can represent a number of different relationships, such as text similarity, topic similarity, or political leaning, the graph must form a structure that relays the likelihood of consumer trust across the sources. 

\begin{figure*}
 \centering
    \includegraphics[width=0.75\textwidth]{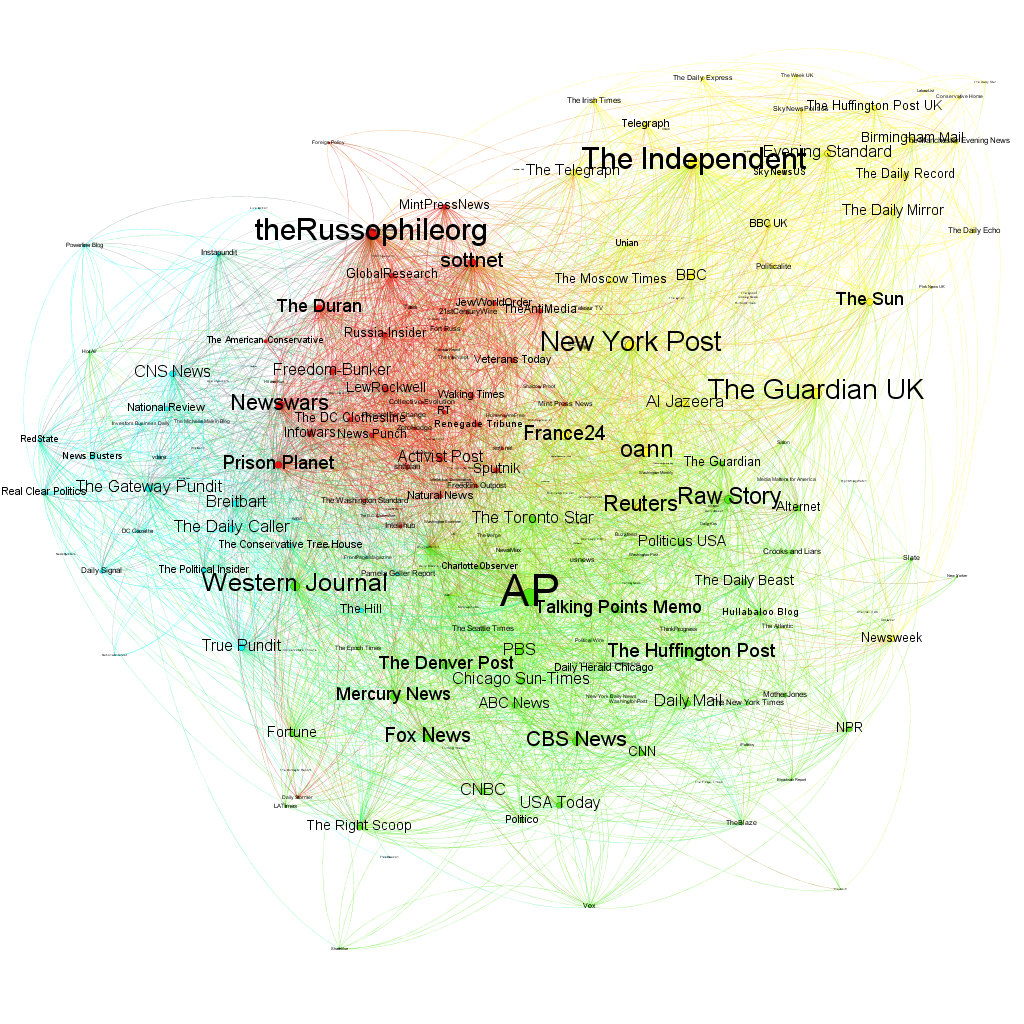}
    \caption{News producer Content Sharing Network (CSN) used as the relationship graph in the \texttt{Trust Nudging} simulation based on data described in Section~\ref{sec:simulation}. Each node represents a news producer, each directed edge represents verbatim copy relationship (where $A \rightarrow B$ means $B$ copies an article from $A$), the size of the node represents the nodes outdegree, and colors represent communities of sources computed using directed modularity.}
    \label{fig:CSN}
\end{figure*}

One example of this structure is a news producer Content Sharing Network (CSN) constructed from a representative sample of news and media sources and articles published them in a similar time frame. Several recent studies build on such datasets have shown that news sources often share (or copy) articles from each other either verbatim or in part~\cite{starbird2018ecosystem, horne2019different, benkler2018network}. When these verbatim copies are formulated as a network, they form meaningful communities of differing sources in the news ecosystem~\cite{horne2019different}, including communities of conspiracy news, partisan news, and mainstream news. Each of these communities represent fairly homogeneous parts of the news ecosystem. Media sources that hold the same ideological values, report on similar events, and write with similar styles are captured in these communities~\cite{horne2019different}. Hence, we expect sources which copy from each other at a high rate to be similar to each other. Extending from this, we expect that sources which are not necessarily directly connected, but near each other in the network space (i.e. same community, peripheral of the same community, etc.) should be alike in some regard. We assume that this similarity leads to a higher probability of trust for these sources than a randomly chosen source of higher quality.  See Figure~\ref{fig:CSN} for an example network.

Using this network we can model a consumer's trust for a source by placing that consumer's reading profile as a node in the network. For example, given a consumer is subscribed to 5 sources on Facebook, we can use these 5 sources to approximate that consumers location in the network space (e.g. by averaging the vector representation of each of the 5 sources in the network using an embedding method) and model trust as the distance of consumer to any source in the network. If we have a measure of quality for each news source (e.g. how many times the source has published false news, etc.), we can ensure our recommendation is of higher quality than the news the consumer currently reads, but also ensure that the consumer has a high chance of trusting that recommendation. Note, depending on the method of approximating the likelihood of trusting a recommendation, the recommended alternatives may not necessarily present very different or diverse points of view, especially if the initial article is on an extreme spectrum of unreliability. However, the main objective is to nudge the user towards higher quality, rather than immediately recommend a gold standard article or source, which may be difficult for the user to accept. We contrast this approach with misinformation detection methods that seek to discredit, demote, or filter out information of lower quality. 

In some sense, this approach can also be contrasted with traditional social recommendation algorithms that aim to only maximize engagement, rather than quality. However, the goals of each recommendation algorithm are fundamentally different. More importantly, inline with the concept of nudging, our approach is providing a recommendation as an alternative, not as the next step. It may be possible to implement both engagement-based recommendations (e.g. Facebook's news feed algorithm or YouTube's up next feature) with the Trust Nudging model, but this is left for future work.

The quality ground truth required by the Trust Nudging model can be built from various criteria, such as whether they have published outright incorrect information in the past, posted corrections to errors, provide transparency of financial interests, or distinguish between news and opinion. There are already organizations that provide this level of analysis, which we will use in our proof-of-concept example~\cite{norregaard2019nela}. While these source-level quality labels are a great start, such measures can be further improved and made more granular.

\section{The Trust Nudging Model in Simulation} \label{sec:simulation}
To illustrate the Trust Nudging method, we construct a proof of concept recommendation system based on real data. We build each of the three sets of information from this data set: the news relationship graph, the source entity, and the user entity. We then make recommendations based on the trust nudge. We show how such a model may work based on simulations based on certain assumptions of user behavior and discuss the results.

\begin{algorithm}
    \SetAlgoLined
    \KwData{$u$=User($S_u$) - the input user profile; \\
    $S$ - the set of sources with quality score $q_s$, leaning $l_s$ and embedded vectors $v_s$;\\ $T$ - the number of iterations T; \\ $L$ - user limited attention; \\ $\alpha$ - cost parameter;}
    \KwResult{Updated user profile $u'$}
    initialization\;
    
    $t \gets 0$\;
    
    \While{$t < T$}{
        $s' = argmin_s(t(s,u)) | q_s > q_u)$\;
        \If {$s'$ is not NULL}
        {
            \If {$|S_u| \geq L$}
            {
                drop\_source$(u,\{S_u,s'\})$\;
            }
            \Else{
                accept\_source$(u,s')$\;
            }
            update\_scores$(u)$\;
        }
    }
    \Return{$u$\;}
    \caption{Simulation algorithm}
    \label{alg:simulation}
\end{algorithm}

\paragraph{\textbf{Data}}
To ground our work on real data, we extract news article data from the NELA-GT-2018 dataset~\cite{norregaard2019nela} and combine it with news article data we collected in the first 9 months of 2019, giving us a total of 1,814,682 articles from 271 sources over 19 months. The NELA-GT-2018 dataset is released with a set of source-level labels from multiple organizations including independent and journalistic venues (NewsGuard, Open Sources, Media Bias/Fact Check, Allsides, BuzzFeed). These labels include both measures of quality and measures of political leaning. We extract and extend these labels for our simulation, discussed below.

\paragraph{\textbf{Relationship graph}}
While the Trust Nudging model can be implemented with many different news relationship graphs, we use a content sharing network (CSN) in this example (as discussed in Section~\ref{sec:usingtrustnudge}). Using our combined dataset, we construct a CSN using the same method used in~\cite{horne2019different}. Specifically, we compute a TF-IDF matrix of all articles in the data set and compute the cosine similarity between each article vector pair (given that each article comes from a different news source). For each pair of article vectors that have a cosine similarity greater or equal to $0.85$, we extract them and order them by the timestamps provided with the data set. This process creates a directed graph $G= (V, E)$, where $V$ is a news source and $E$ is a directed weighted edge representing articles shared. Edges are directed towards publishers that copy articles (inferred by the timestamps). We normalize the weight of each edge in the network by the number of articles published in total by the source. The end result is a near-verbatim content sharing network of news sources. This network can be found in Figure~\ref{fig:CSN}. After constructing the network (which naturally filters down both articles and sources), we have 102,879 pairs of articles and 195 sources (which have copied or been copied from) over 19 months. The constructed network can be found in Figure~\ref{fig:CSN}. 

\paragraph{\textbf{Source entity}}
In order build our simulation, we must know both the approximate quality of a news source and the political leaning. Both quality and political leaning can be extracted from the ground truth provided in the NELA-GT-2018 dataset. Specifically, we determine the \textbf{quality score} of a source by normalizing the scores provided by NewsGuard. In addition, we give a score of 0 to sources that have been flagged with at least one of the following labels: by Open sources as \emph{fake}, \emph{conspiracy}, \emph{junksci}, or \emph{unreliable} or by Media Bias/Fact Check as \emph{conspiracy}, \emph{pseudoscience}, or \emph{questionable source}. The \textbf{leaning score} of a source is computed by averaging the scores from fields AllSides \emph{bias rating}, Buzzfeed \emph{leaning}, Media Bias/Fact Check \emph{left bias} or \emph{right bias}, the resulting leaning score is normalized to the interval $[-1,1]$, where $-1$ indicates a left-wing bias and $1$ a right-wing bias. For the few sources that do not have any labeled data in NELA-GT-2018, we estimate the quality score and leaning by averaging its neighbors quality score and leaning in the CSN. It can be argued that this may be a noisy approximation of both quality and leaning, but for the purposes of our simulated example these potentially noisy labels are fine.
More specifically, for $s$ be a source in $S$, we define the \textbf{quality score} $q_s$ of $s$ as a number in $[0,1]$ obtained via the aforementioned approach. Similarly, the \textbf{leaning} $l_s$ of $s$ is a number in $[-1,1]$. 

Additionally, we obtain a vector based representation $v_s$ for each source $s$ by embedding the $s$ is a vector derived from the embedding CSN using node2vec \cite{grover2016node2vec} with 64-dimensional vectors. This representation captures the closeness, or similarity, of nodes in the CSN.

\paragraph{\textbf{User entity}}
Lastly, we need a set of users to nudge in the simulation. Let $u$ user in the set of users $U$. User $u$ has a set $S_u$ of trusted sources, which has a maximum size specified by a \textbf{limited attention} parameter $L$. We assume that a user has a limited number of sources that it can attend to, or trust, at any moment. From set $S_u$ we determine the quality score, leaning, and CSN representation, $q_u$, $l_u$, $v_u$, respectively, associated with $u$ on their news consumption profile by averaging the quality score, leaning, and CSN vectors of the sources in $S_u$.

\paragraph{\textbf{Trusted recommendation}}
Now that we have a news relationship graph, a set of sources with ground truth, and a set of users with reading profiles, we can begin making recommendations. A recommendation consists of suggesting a new source $s'$ to user $u$ at a time $t$. The probability that user $u$ will trust $s'$ depends on $u$'s profile. We denote it by the conditional probability $p_u(s'|S_u)$. Our model considers both the leaning and the CSN representation of $u$ and $s'$ to compute $p_u(s'|S_u)$. More specifically, we compute the leaning differential $\Delta l_{u,s'}$ as the normalized distance between $l_u$ and $l_{s'}$, and we compute the source distance $d_{us'}$ as the cosine distance between $v_u$ and $v_{s'}$. We define the \textbf{trust cost} of recommending $s'$ to $u$ as:

\begin{align}
    \label{eq:trust_cost}
    t(s',u) = p_u(s'|S_u) = (1-\alpha) \Delta l + \alpha d_{us'}
\end{align}

Where $\alpha \in (0,1)$ is a hyper-parameter to control the weight of the source distance and the leaning in the function, $\Delta l = \frac{|{l_u - l_{s'}|}}{2}$, and $d_{us'} = 1- \cos(v_u,v_{s'})$. Concretely, we define the trust cost as a function of how dissimilar the new source is from the user's profile with respect to their position in the CSN and their political leaning.

If the user accepts the recommendation, source $s'$ is added to $S_u$. If $|S_u| = L$, the user drops one source at random chosen from the following distribution:

\begin{align*}
    p_{drop}(s) = \frac{t(s, u)}{\sum t(s_i,u)} & \text{for } s_i \in S_u \cup \{s'\} 
\end{align*}

Note that $s'$ may be dropped in this process, which means it was rejected by the user. If $s'$ is not dropped, the user leaning and quality score are updated.

\paragraph{\textbf{Simulation of trusted recommendations over time}}
Given a user profile $u$, the simulation runs for $T$ discrete time steps. At each time step $t \leq T$, the model produces a recommended source $s'$ meeting the criteria that $q_{s'} > q_u$ requiring the least trust cost, thus:

\begin{align*}
    s' = argmin_s(t(s,u)|q_s > q_u)
\end{align*}

If the user's set of sources $|S_u| \geq L$, one source in $S_u \cup {s'}$ is chosen and discarded (this can be thought of as the source being distrusted by the user or simply no longer being read by the user due to limited attention). Otherwise, $s'$ is accepted with probability $1-t(s',u)$. If $s'$ is accepted, $u$'s profile is updated by calculating the new means for $q_u$, $l_u$, and $v_u$. This procedure is repeated for $T$ iterations, when the updated profile of $u$ is obtained. Notice that once $q_u=1$, no recommendation is made since the user has reached the maximum score possible according to the model, we refer to the earliest $t$ at which $q_u=1$ as the \textbf{convergence point}. Algorithm \ref{alg:simulation} outlines the simulation procedure.

\begin{table*}[h]
    \centering
    \begin{tabular}{c|c|c|c}
    \textbf{User} & \textbf{Starting Quality} & \textbf{Starting Leaning} & \textbf{Starting Sources}\\
    \hline
     User A & 0.075 & 0.622 & Infowars, Newswars, Prison Planet, Veterans Today, Natural News\\
     User B & 0.350 & -0.854 & Daily Kos, Shareblue, Bipartisan Report, Delaware Liberal, Addicting Info\\
     User C & 0.524 & 1.0 & Breitbart, Conservative TreeHouse, CNS News, The Epoch Times, Western Journal\\
     User D & 0.098 & -0.158 & 21st Century Wire, Mint Press News, Global Research, The Duran, Intellihub\\
     & & &
    \end{tabular}
    \caption{Starting User Profiles used in simulation. Quality score is a number between 0 and 1, where 1 is the highest quality. For each user this is computed by taking the average quality score of the news sources they read. Leaning is a number between -1 and 1, where -1 is far left leaning and 1 is far right leaning. Again, this score is computed by averaging the political leaning of the users reading profile.}
    \label{tab:userprofile}
\end{table*}

\begin{table*}[h]
    \centering
    \begin{tabular}{c|c|c|c|c}
    \textbf{User} & \textbf{Ending Quality} & \textbf{Ending Leaning} & \textbf{Ending Sources} & \textbf{Time}\\
    \hline
     User A & 1.0 & 0.516 & National Review, Real Clear Politics, The American Conservative, Politico, Fortune & 54\\
     User B & 1.0 & -0.500 & CBS News, The New York Times, The Guardian, Washington Post, The Denver Post & 14\\
     User C & 1.0 & 0.516 & Real Clear Politics, National Review, Fortune, The American Conservative, Politico & 20\\
     User D & 1.0 & -0.250 & FiveThirtyEight, Business Insider, NPR, The Hill, BBC & 59\\
     & & & &
    \end{tabular}
    \caption{User Profiles After simulation. Quality score is a number between 0 and 1, where 1 is the highest quality. For each user this is computed by taking the average quality score of the news sources they read. Leaning is a number between -1 and 1, where -1 is far left leaning and 1 is far right leaning. Again, this score is computed by averaging the political leaning of the users reading profile. Time is the number of iterations the user took to get to a quality of 1.0.}
    \label{tab:userend}
\end{table*}

\newcommand{\figwidth}{0.40}
\begin{figure*}
\centering

    \begin{subfigure}{\figwidth\textwidth}
    \centering
    \textbf{{\large Quality Score}} 
    \end{subfigure}
    \begin{subfigure}{\figwidth\textwidth}
    \centering
    \textbf{{\large Political Leaning}}
    \end{subfigure}

    \textbf{\large User A}\begin{subfigure}{\figwidth\textwidth}
        \includegraphics[width=\textwidth]{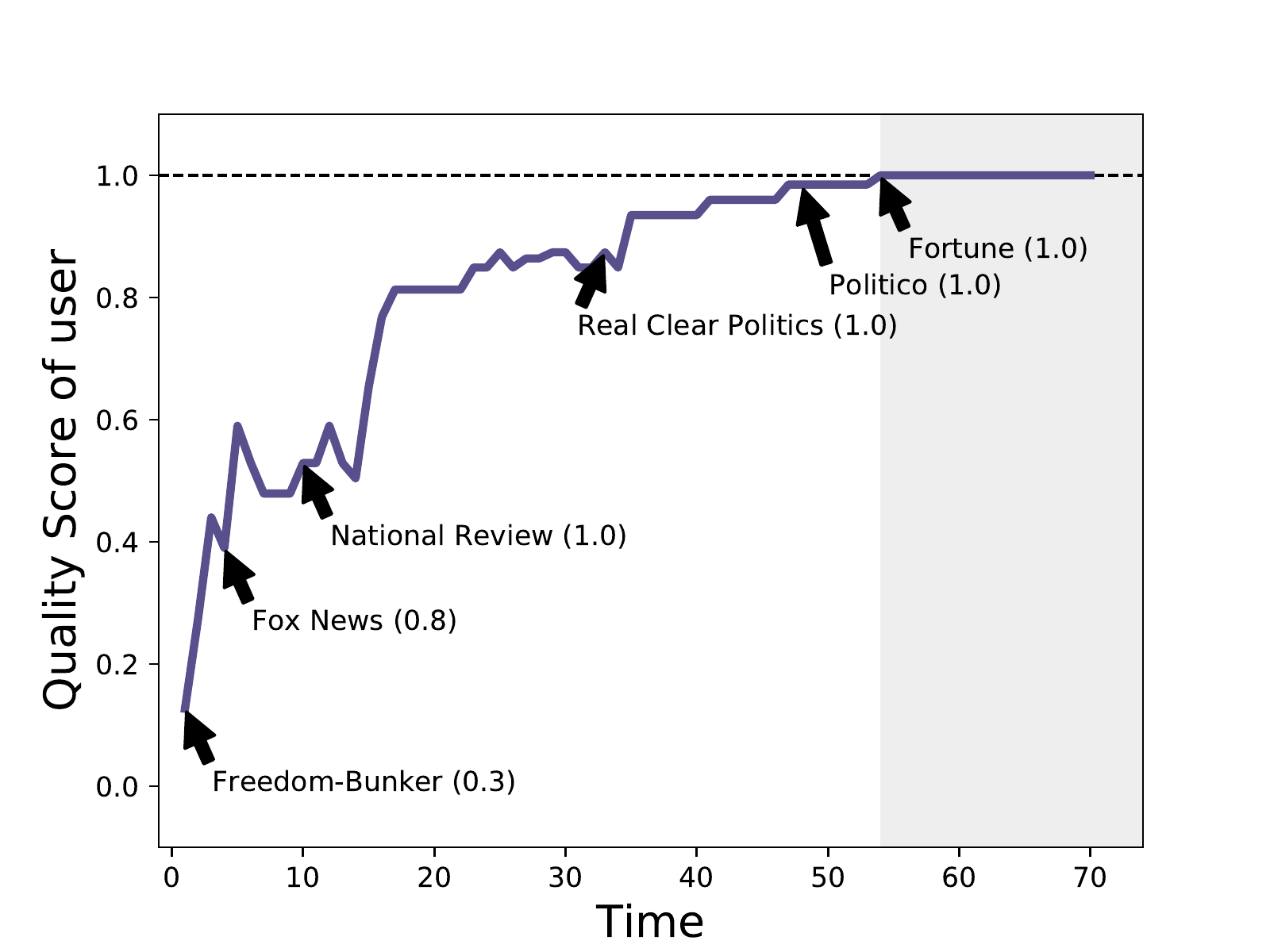}
    \end{subfigure}
    \begin{subfigure}{\figwidth\textwidth}
        \includegraphics[width=\textwidth]{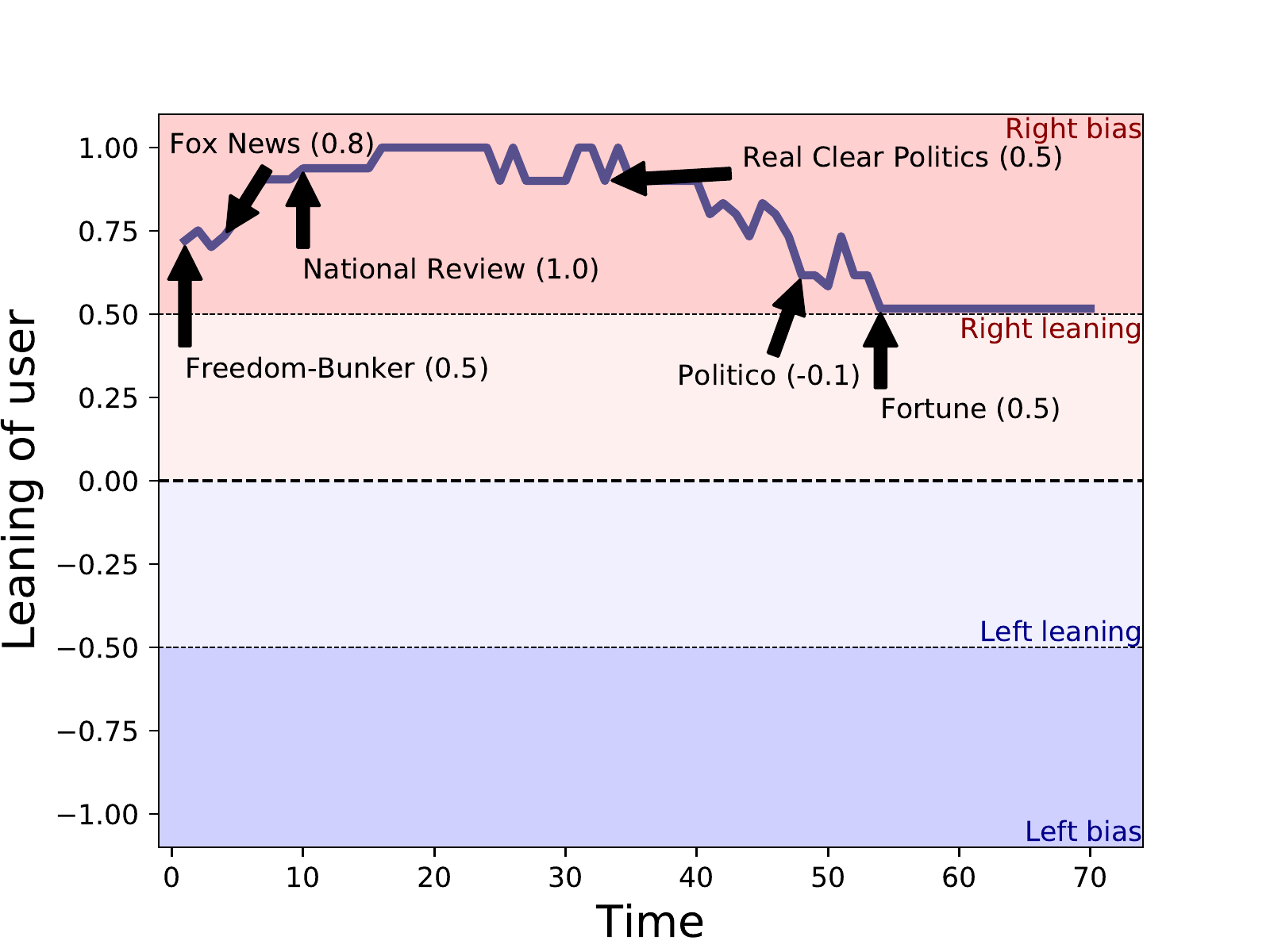}
    \end{subfigure}
    
    \textbf{\large User B}\begin{subfigure}{\figwidth\textwidth}
        \includegraphics[width=\textwidth]{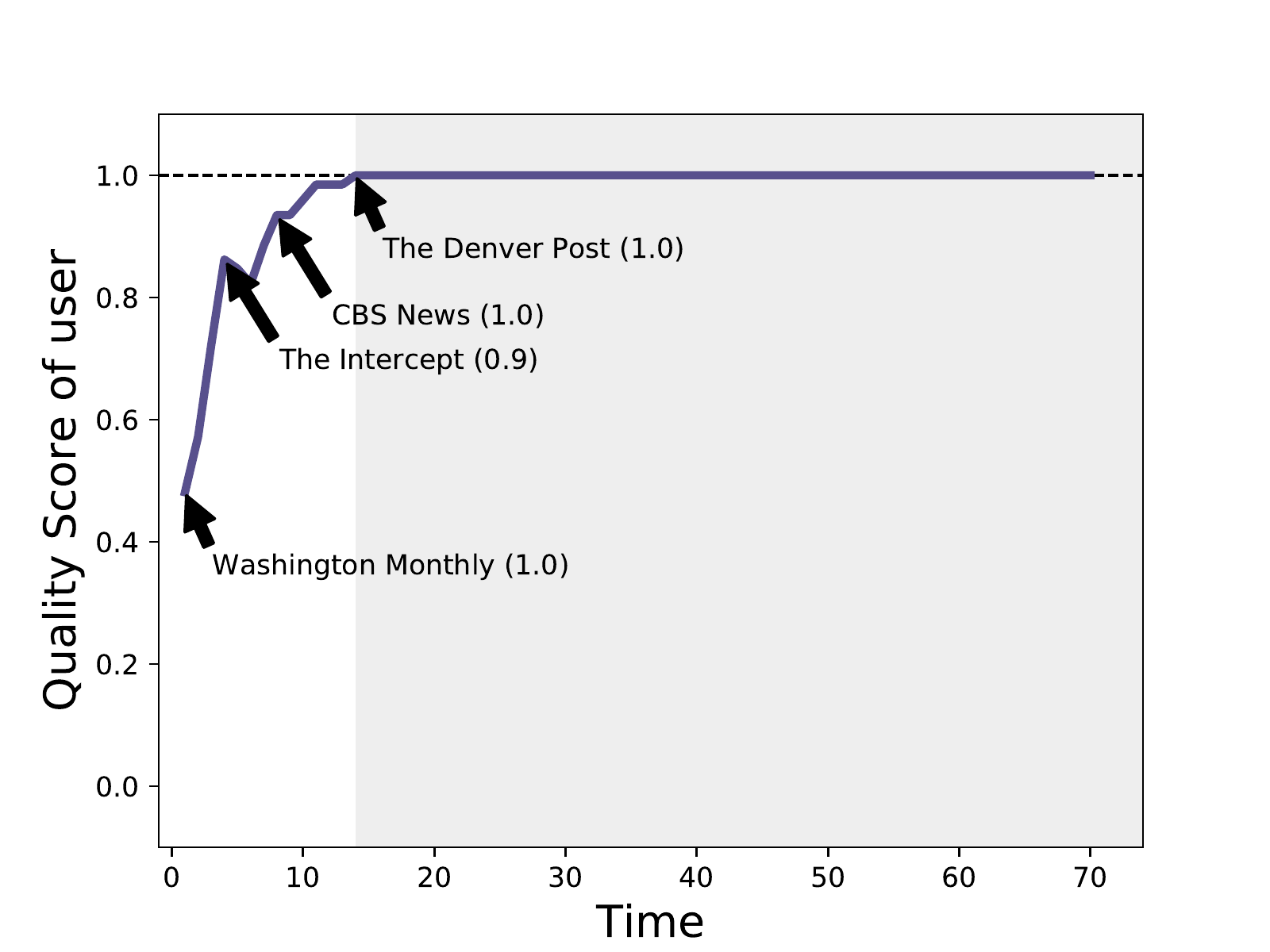}
    \end{subfigure}
    \begin{subfigure}{\figwidth\textwidth}
        \includegraphics[width=\textwidth]{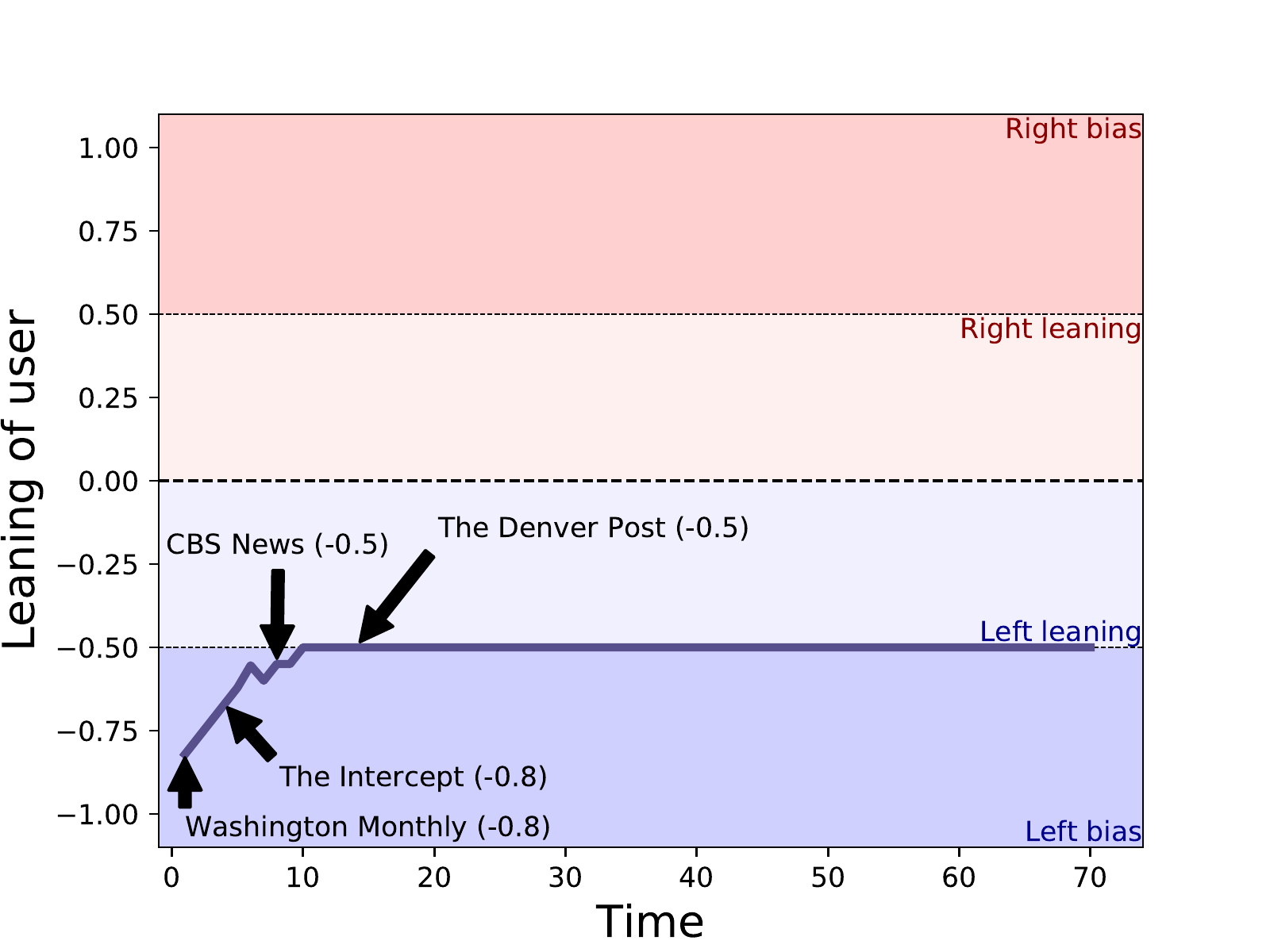}
    \end{subfigure}
    
    \textbf{\large User C}\begin{subfigure}{\figwidth\textwidth}
        \includegraphics[width=\textwidth]{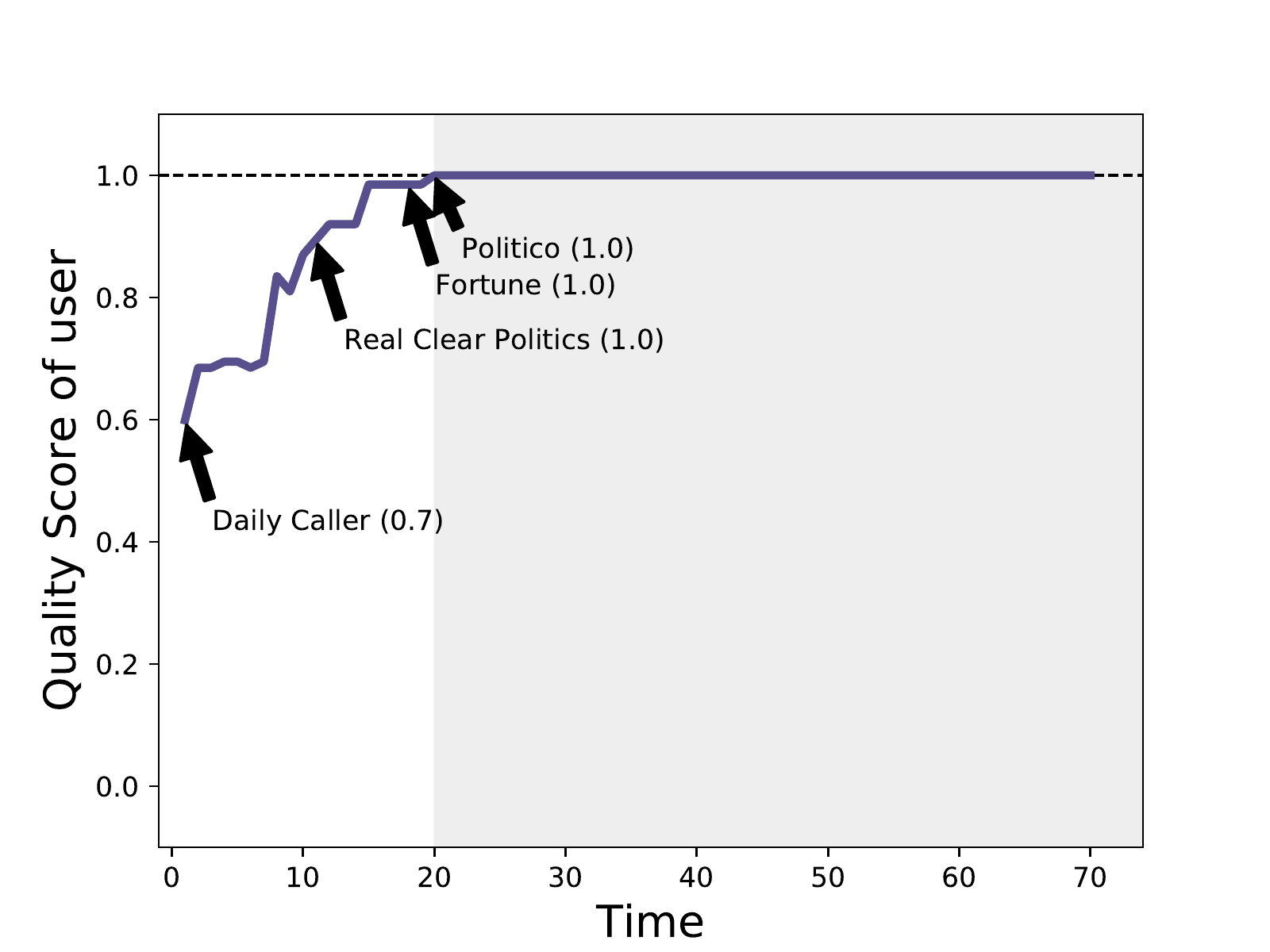}
    \end{subfigure}
    \begin{subfigure}{\figwidth\textwidth}
        \includegraphics[width=\textwidth]{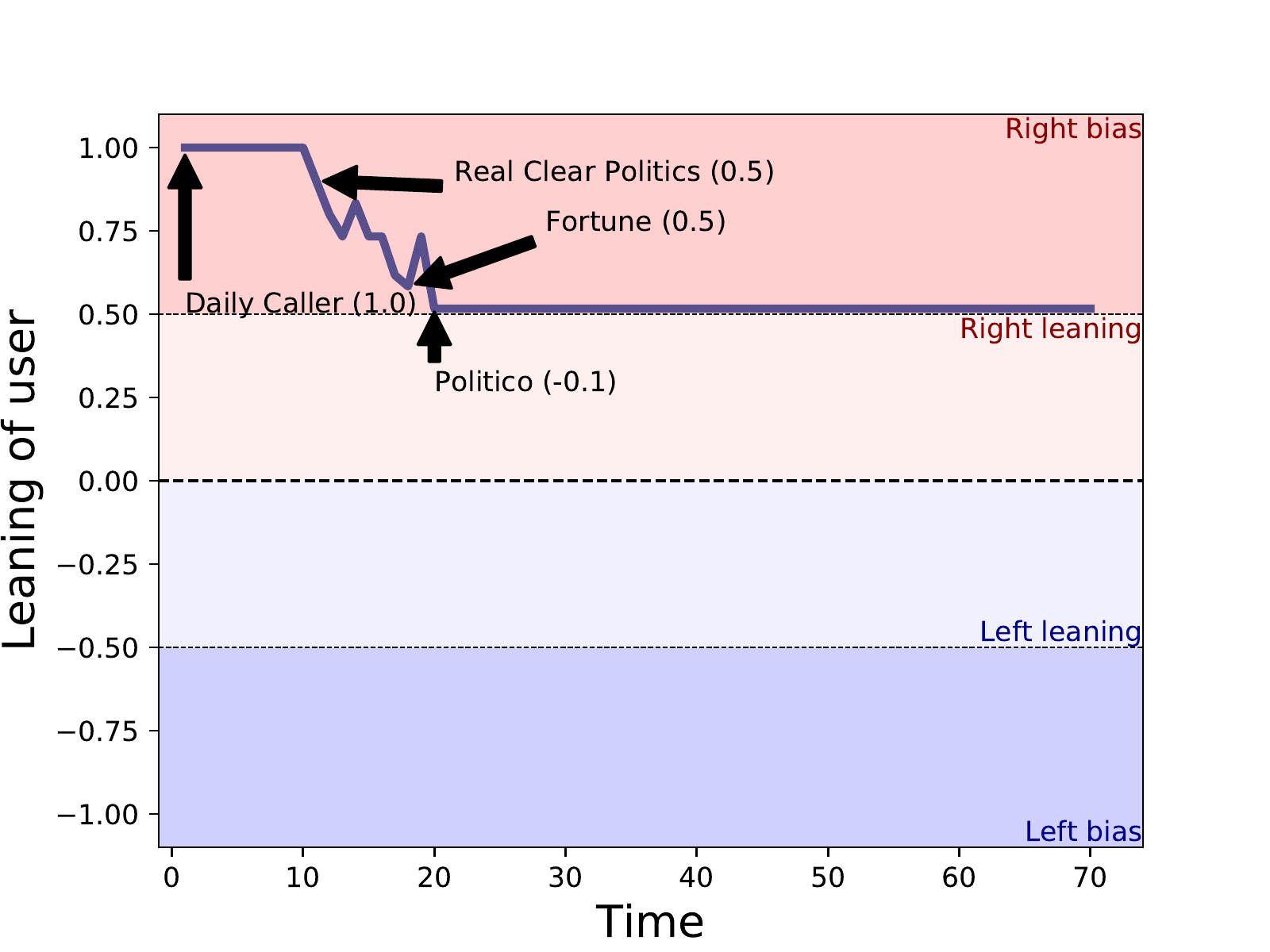}
    \end{subfigure}
    
    \textbf{\large User D}\begin{subfigure}{\figwidth\textwidth} 
        \includegraphics[width=\textwidth]{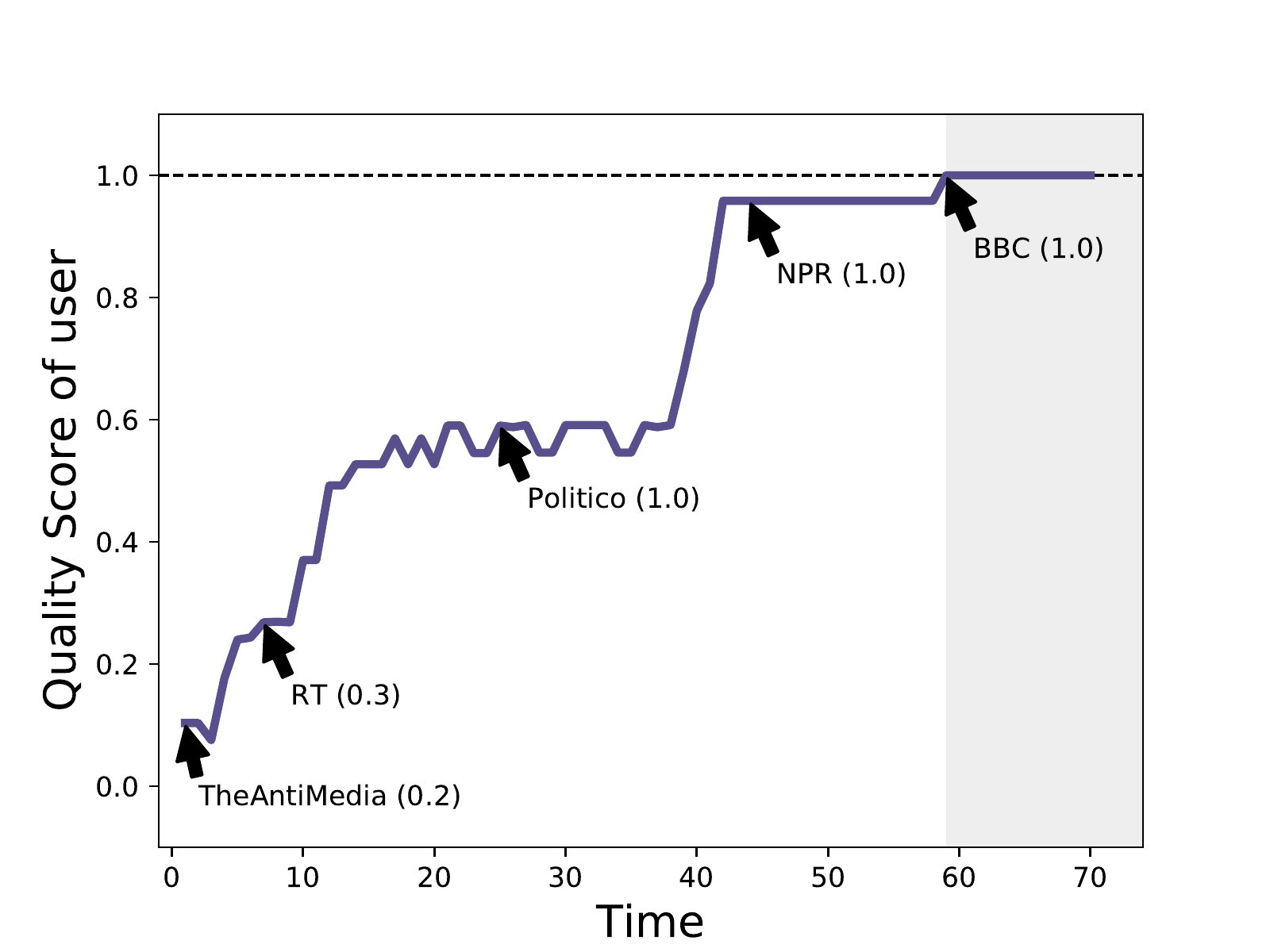}
    \end{subfigure}
    \begin{subfigure}{\figwidth\textwidth}
        \includegraphics[width=\textwidth]{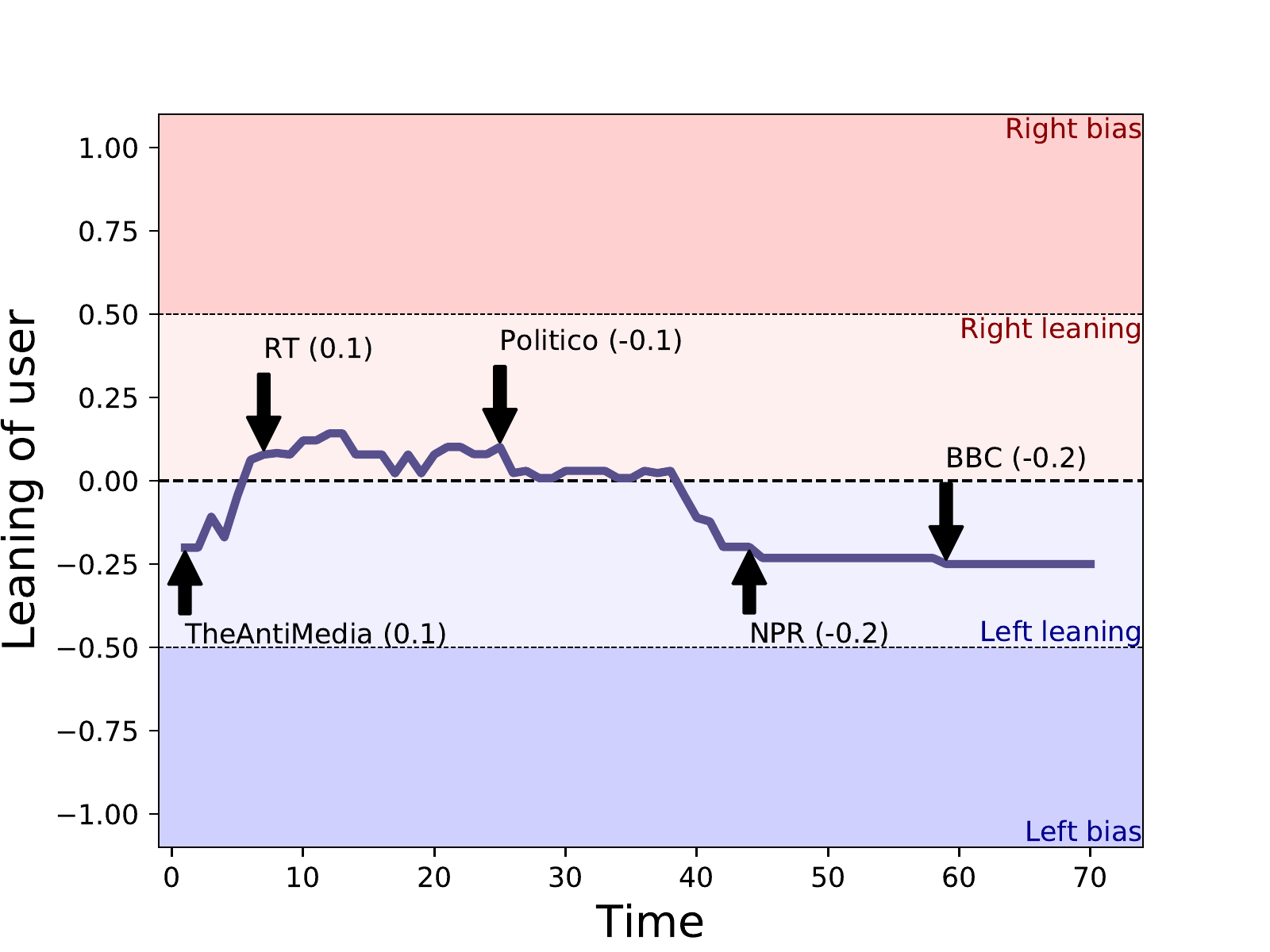}
    \end{subfigure}
    \caption{Recommendation pathways. On the left are the trajectories of the quality score of the user, some sources are emphasized along the pathway, enclosed in parentheses are the quality score of each source; the gray box indicates convergence. On the right are the leaning trajectories, enclosed in parentheses are the leaning score of each source.}
    \label{fig:pathways}
\end{figure*}

\subsection{Simulation Results}
We ran the simulation on four synthetically developed user profiles. The profiles were chosen to depict users with distinct characteristics with respect to political leaning and the types of sources they trust (conspiracy, left-leaning, right-leaning, etc.). User profiles are shown in Table \ref{tab:userprofile}.

\paragraph{\textbf{Pathways to high quality news}}
Figure \ref{fig:pathways} shows the quality score (left side) and leaning (right side) trajectories for each user. All users converge to a quality of 1, although at different times, the gray box indicates convergence to a quality of 1. Some of the sources along each pathway are emphasized, their scores (quality and leaning) are annotated in parentheses in each figure. The ending profile and scores of users can be seen in Table \ref{tab:userend}. Observing the simulation results, we are not only able to see the final set of sources of users, but also the pathway taken by each of them in order to reach a high quality news consumption standard at each time step. 

Importantly, the first recommendation made for each user is of higher quality than the user's average reading profile, but is very similar to both their reading profile in the CSN space and their reading profile in terms of the political leaning. While the quality is always higher than the user's average reading quality, it may not be much higher depending on the extremity of leaning and connectedness of the reading profile in the CSN. For example, User A is first recommended Freedom Bunker, which is a low quality (quality score of 0.3) right-wing (leaning score of 0.5) source, but this source is both higher quality (user's quality score of 0.075) and less extreme (user's leaning score of 0.622) than the user's reading profile. User A had a 96.7\% chance of accepting that recommendation. Similarly, User D is recommended a low quality (quality score of 0.2) left-leaning (leaning score of 0.1) source that is of higher quality than the user's reading profile. User D had a 94.2\% chance of accepting this recommendation. In both cases, the user's overall quality score improves after accepting the first recommendation. As expected, user's with less extreme profiles have a higher chance of accepting the first recommendation and converge to high quality quicker. User B had a 97.8\% chance of accepting the first recommendation and User C had a 99.3\% chance of accepting the first recommendation.

As time progresses, users' likelihood of trusting a recommended source will remain small. However, as the users accept recommended sources, their reading profile will slowly change, allowing for higher quality (or more dissimilar to the original reading profile) sources to be recommended. It is clear that the jump from Freedom Bunker to Fox News is less costly than the jump from Freedom Bunker to Politico. Similarly, the jump from TheAntiMedia to RT is less costly than the jump from TheAntiMedia to NPR. 

Interestingly, as a side effect of increasing quality, the model also lessens the partisan extremity of the news being read (with the exception of User D, who already was not very extreme in political leaning, only quality). 



\begin{figure}
    \centering
    \includegraphics[width=\figwidth\textwidth]{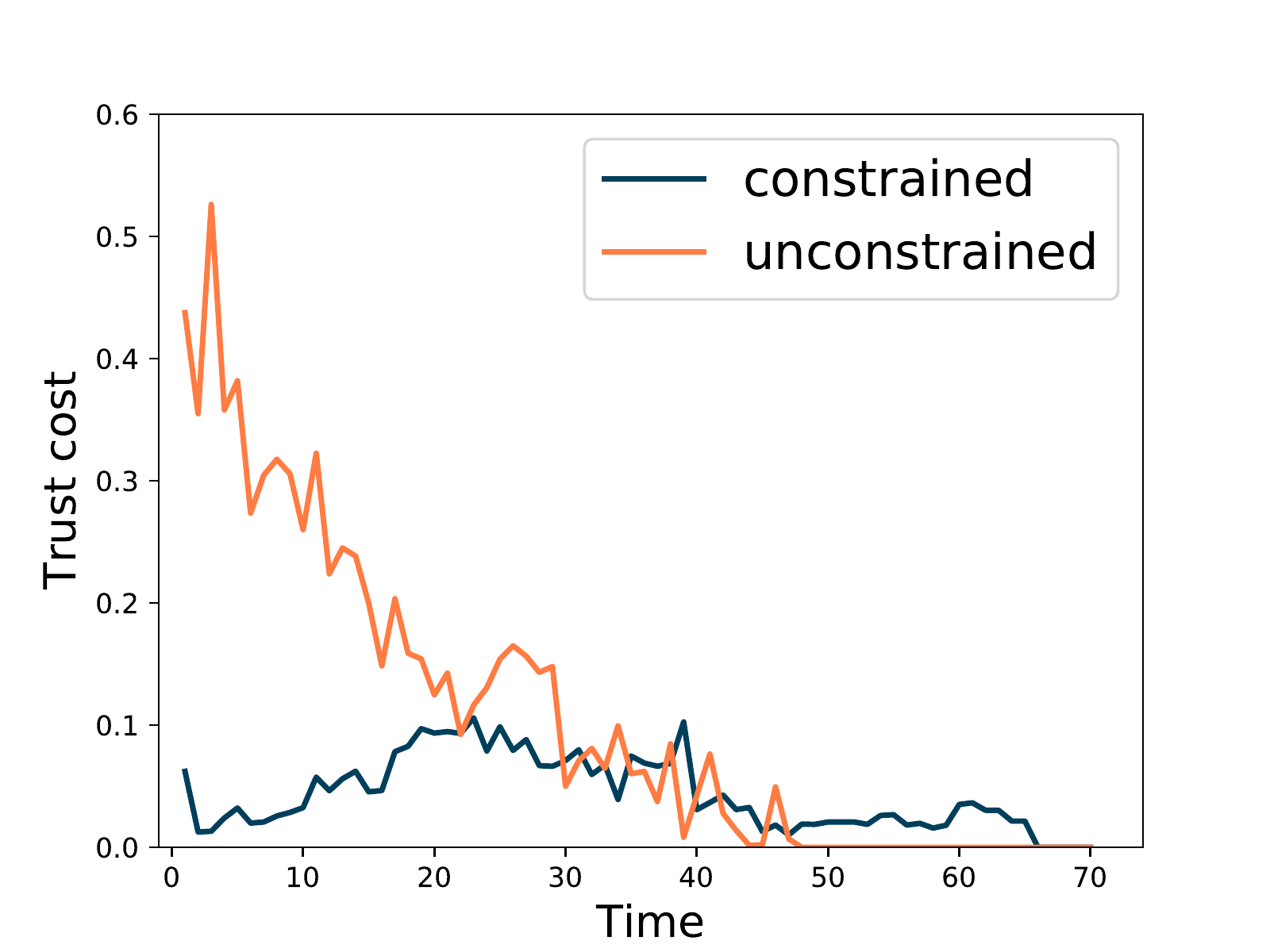}
    \caption{Comparison of the trust cost for the constrained and the unconstrained models for User A. Notice the initially trust cost for the unconstrained model. This suggests that users are unlikely to accept the recommendations at the beginning.}
    \label{fig:trust_cost}
\end{figure}

\paragraph{\textbf{Comparison to a trust-unconstrained model}}
The Trust Nudging model is a trust-constrained model because the choice of recommendation is a function of the trust cost. A trust-unconstrained model would provide recommendations that are not necessarily functions of trust, or do not take trust directly into account when making decisions. A trust-unconstrained model can be thought of as a recommendation system that recommends gold standard news no matter the consumer. For comparison, we modified our model to become trust-unconstrained by removing the trust function from the source selection step and simply picking a source that increments the overall quality score of the user. We then looked at the progression of the trust cost for each iterations step $t$ comparing the constrained and unconstrained models. As seen in Figure \ref{fig:trust_cost}, the trust cost for the unconstrained model started at a much higher value than the constrained one, as time passed, the trust cost started to reduce because the user had incorporated high quality source to their profile and became naturally closer to good sources in the CSN space. Even though both models converge at approximately the same time, one may argue that it is unlikely for a user to accept drastic changes during the first recommendations since they are highly unlikely to trust them. In the simulation model, the user is repeatedly exposed to the recommendation, eventually leading to acceptance. However, in real life, a user may begin to distrust the recommendation system or stop using the platform due to repeated and abrupt recommendations.



\section{Discussion and Future Work}
In this paper, we proposed a trust-based news recommendation model which nudges consumers towards higher quality without demanding a substantial challenge of one's beliefs. The potential benefit of such a model is the ability to move extreme or conspiracy news consumers towards higher quality information, a population which is hard to persuade. We provide a proof of concept for the Trust Nudging model through simulation. Using this simulation we show that the model can slowly improve news consumption over time while recommending sources which the user is likely to trust. We also show that, as a side effect of improved quality, hyper-partisan news consumers eventually consume less partisan news. Although conclusions that can be drawn from this simulation are limited, the purpose of this paper is to open up a new technical research path for misinformation mitigation, namely using nudges in news algorithms rather than filtering, demoting, and discrediting. 

There are many ways this model can be improved. For example, rather than only defining trust as a function of similarity in the CSN, more granular relationship graphs can be used, such as the similarity of writing style between sources. We can also model trust as a function of the difference in attitude, or stance, towards a topic, provided that certain sources are more reliable or more similar when reporting on specific topics. Additionally, multiple news relationship graphs can be used at once through a multi-layer network. The value of these various news graphs, in terms of representing consumer trust, should be tested.

This model can also be improved through more complex objective functions. For example, rather than only operating on quality, the model can also nudge towards view-point diversity. Such a system may expose users to better or more diverse information even if it is not clicked. Previous research shows that titles play a big role in disseminating misinformation. Many readers
will consume news passively and form opinions based on the titles
alone. In such cases, even the exposure to slightly different framing of information may create some amount of doubt in low quality information, especially since sources are chosen based on potentially trusted sources. Moreover, this type of model can create financial incentive for sources to provide a better quality of information instead of models based on engagement which tend to favor more sensational information and models based on gold standards that can be hard to define in some cases. In an effective implementation of our method, many sources have a chance to get engagement by providing better quality information.

Of course there are many unknowns about this type of nudging recommendation model that should be assessed. While in theory our simulated trust calculation makes sense, in actuality trust is likely much more complex. Other trust factors, such as user stances on specific topics, whether a friend had read or shared the news, or coherence of news story, could be accounted for, as long a quality measure is still held as the objective. It is likely these more granular trust factors are consumer-dependent and can change across readers. Namely, a user's trust may be modeled not only with respect to their news consumption profile, but also in terms of how much influence they receive from their peers and their susceptibility to such, thus adding new dimensions to the definition of trust. This increased trust complexity may produce different behavior over time. For example, rather than eventually converging to the highest quality sources, there may be a point in which consumers can no longer be nudged, hence plateauing the quality we can achieve. User studies should be implemented to assess interaction with the algorithm. 

There are also some potentially negative consequences of using this type of model, as news consumers will continue to
be exposed to bad and incorrect information without seeing any
warnings regarding its incorrectness. Hence, users will continue to form
incorrect beliefs that may be very hard to overcome. In addition, even
if higher quality information is available, users may never choose
to read it as they do not find it engaging. The hope is that this exposure to low veracity news will diminish as the consumers are nudged; however, it is unknown how long it will take consumers to gain trust for higher quality sources and when/if they will stop accepting nudges. Another potentially negative consequence is the newly created incentive for malicious sources to game the system. It is possible that a smart malicious source can work to become a 'middle quality' source, which will be recommended to users who previously consumed extreme news. These malicious sources may even be able to control both the extreme news being consumed and the middle quality news being recommended. The robustness to adversarial manipulation should be explicitly tested. Finally, questions of how this model can work in harmony with other policies and community standards regarding unacceptable content requires additional research.

\bibliographystyle{IEEEtran}
\bibliography{references, exported_references}

\begin{thebibliography}{10}
\providecommand{\url}[1]{#1}
\csname url@samestyle\endcsname
\providecommand{\newblock}{\relax}
\providecommand{\bibinfo}[2]{#2}
\providecommand{\BIBentrySTDinterwordspacing}{\spaceskip=0pt\relax}
\providecommand{\BIBentryALTinterwordstretchfactor}{4}
\providecommand{\BIBentryALTinterwordspacing}{\spaceskip=\fontdimen2\font plus
\BIBentryALTinterwordstretchfactor\fontdimen3\font minus
  \fontdimen4\font\relax}
\providecommand{\BIBforeignlanguage}[2]{{%
\expandafter\ifx\csname l@#1\endcsname\relax
\typeout{** WARNING: IEEEtran.bst: No hyphenation pattern has been}%
\typeout{** loaded for the language `#1'. Using the pattern for}%
\typeout{** the default language instead.}%
\else
\language=\csname l@#1\endcsname
\fi
#2}}
\providecommand{\BIBdecl}{\relax}
\BIBdecl

\bibitem{popat2016credibility}
K.~Popat, S.~Mukherjee, J.~Str{\"o}tgen, and G.~Weikum, ``Credibility
  assessment of textual claims on the web,'' in \emph{Proceedings of ACM
  CIKM}.\hskip 1em plus 0.5em minus 0.4em\relax ACM, 2016, pp. 2173--2178.

\bibitem{hassan2017claimbuster}
N.~Hassan, G.~Zhang, F.~Arslan, J.~Caraballo, D.~Jimenez, S.~Gawsane, S.~Hasan,
  M.~Joseph, A.~Kulkarni, A.~K. Nayak \emph{et~al.}, ``Claimbuster: the
  first-ever end-to-end fact-checking system,'' \emph{Proceedings of the VLDB
  Endowment}, vol.~10, no.~12, pp. 1945--1948, 2017.

\bibitem{barron2019proppy}
A.~Barr{\'o}n-Cedeno, I.~Jaradat, G.~Da~San~Martino, and P.~Nakov, ``Proppy:
  Organizing the news based on their propagandistic content,''
  \emph{Information Processing \& Management}, 2019.

\bibitem{horne2018accessing}
B.~D. Horne, W.~Dron, S.~Khedr, and S.~Adal{\i}, ``Assessing the news
  landscape: A multi-module toolkit for evaluating the credibility of news,''
  in \emph{WWW Companion}, 2018.

\bibitem{baly2018}
R.~Baly, G.~Karadzhov, D.~Alexandrov, J.~Glass, and P.~Nakov, ``Predicting
  factuality of reporting and bias of news media sources,'' in
  \emph{Proceedings of 2018 EMNLP}, 2018.

\bibitem{starbird2018ecosystem}
K.~Starbird, A.~Arif, T.~Wilson, K.~Van~Koevering, K.~Yefimova, and
  D.~Scarnecchia, ``Ecosystem or echo-system? exploring content sharing across
  alternative media domains,'' 2018.

\bibitem{starbird2019disinformation1}
K.~STARBIRD, A.~ARIF, and T.~WILSON, ``Disinformation as collaborative work:
  Surfacing the participatory nature of strategic information operations,''
  2019.

\bibitem{amazeen2015sometimes}
M.~Amazeen, ``Sometimes political fact-checking works. sometimes it doesn’t.
  here’s what can make the difference,'' \emph{The Washington Post}, vol.~3,
  2015.

\bibitem{sunstein2009conspiracy}
C.~R. Sunstein and A.~Vermeule, ``Conspiracy theories: Causes and cures,''
  \emph{Journal of Political Philosophy}, vol.~17, no.~2, pp. 202--227, 2009.

\bibitem{zannettou2018gab}
S.~Zannettou, B.~Bradlyn, E.~De~Cristofaro, H.~Kwak, M.~Sirivianos,
  G.~Stringini, and J.~Blackburn, ``What is gab: A bastion of free speech or an
  alt-right echo chamber,'' in \emph{Companion Proceedings of the The Web
  Conference 2018}, 2018, pp. 1007--1014.

\bibitem{turcotte2015news}
J.~Turcotte, C.~York, J.~Irving, R.~M. Scholl, and R.~J. Pingree, ``News
  recommendations from social media opinion leaders: Effects on media trust and
  information seeking,'' \emph{Journal of Computer-Mediated Communication},
  vol.~20, no.~5, pp. 520--535, 2015.

\bibitem{hanitzsch2018caught}
T.~Hanitzsch, A.~Van~Dalen, and N.~Steindl, ``Caught in the nexus: A
  comparative and longitudinal analysis of public trust in the press,''
  \emph{The International Journal of Press/Politics}, vol.~23, no.~1, pp.
  3--23, 2018.

\bibitem{thaler2009nudge}
R.~H. Thaler and C.~R. Sunstein, \emph{Nudge: Improving decisions about health,
  wealth, and happiness}.\hskip 1em plus 0.5em minus 0.4em\relax Penguin, 2009.

\bibitem{guacho2018semi}
G.~B. Guacho, S.~Abdali, N.~Shah, and E.~E. Papalexakis, ``Semi-supervised
  content-based detection of misinformation via tensor embeddings,'' in
  \emph{2018 ASONAM}.\hskip 1em plus 0.5em minus 0.4em\relax IEEE, 2018, pp.
  322--325.

\bibitem{ruchansky2017csi}
N.~Ruchansky, S.~Seo, and Y.~Liu, ``Csi: A hybrid deep model for fake news
  detection,'' in \emph{Proceedings of the 2017 ACM on Conference on
  Information and Knowledge Management}.\hskip 1em plus 0.5em minus 0.4em\relax
  ACM, 2017, pp. 797--806.

\bibitem{horne2019robust}
B.~D. Horne, J.~Nørregaard, and S.~Adal{\i}, ``Robust fake news detection over
  time and attack,'' \emph{ACM Transactions of Intelligent Systems Technology},
  2019.

\bibitem{nyhan2010corrections}
B.~Nyhan and J.~Reifler, ``When corrections fail: The persistence of political
  misperceptions,'' \emph{Political Behavior}, vol.~32, no.~2, pp. 303--330,
  2010.

\bibitem{wood2016elusive}
T.~Wood and E.~Porter, ``The elusive backfire effect: mass attitudes steadfast
  factual adherence,'' \emph{Political Behavior}, pp. 1--29, 2016.

\bibitem{johnson1994sources}
H.~M. Johnson and C.~M. Seifert, ``Sources of the continued influence effect:
  When misinformation in memory affects later inferences.'' \emph{Journal of
  Experimental Psychology: Learning, Memory, and Cognition}, vol.~20, no.~6, p.
  1420, 1994.

\bibitem{ecker2017reminders}
U.~K. Ecker, J.~L. Hogan, and S.~Lewandowsky, ``Reminders and repetition of
  misinformation: Helping or hindering its retraction?'' \emph{Journal of
  Applied Research in Memory and Cognition}, vol.~6, no.~2, pp. 185--192, 2017.

\bibitem{swire2019they}
B.~Swire-Thompson, U.~K. Ecker, S.~Lewandowsky, and A.~J. Berinsky, ``They
  might be a liar but they’re my liar: Source evaluation and the prevalence
  of misinformation,'' \emph{Political Psychology}, 2019.

\bibitem{thorson2016belief}
E.~Thorson, ``Belief echoes: The persistent effects of corrected
  misinformation,'' \emph{Political Communication}, vol.~33, no.~3, pp.
  460--480, 2016.

\bibitem{rich2016continued}
P.~R. Rich and M.~S. Zaragoza, ``The continued influence of implied and
  explicitly stated misinformation in news reports.'' \emph{Journal of
  experimental psychology: learning, memory, and cognition}, vol.~42, no.~1,
  p.~62, 2016.

\bibitem{garrett2013undermining}
R.~K. Garrett, E.~C. Nisbet, and E.~K. Lynch, ``Undermining the corrective
  effects of media-based political fact checking? the role of contextual cues
  and na{\"\i}ve theory,'' \emph{Journal of Communication}, vol.~63, no.~4, pp.
  617--637, 2013.

\bibitem{Adali:2013}
S.~Adal{\i}, \emph{Modeling Trust Context in Networks}.\hskip 1em plus 0.5em
  minus 0.4em\relax Springer Briefs, 2013.

\bibitem{Hilligoss:2008}
B.~Hilligoss and S.~Y. Rieh, ``Developing a unifying framework of credibility
  assessment: Construct, heuristics and interaction in context,''
  \emph{Information Processing and Management}, vol.~44, pp. 1467--1484, 2008.

\bibitem{lewandowsky2012misinformation}
S.~Lewandowsky, U.~K. Ecker, C.~M. Seifert, N.~Schwarz, and J.~Cook,
  ``Misinformation and its correction: Continued influence and successful
  debiasing,'' \emph{Psychological Science in the Public Interest}, vol.~13,
  no.~3, 2012.

\bibitem{horneadali2017}
B.~Horne and S.~Adali, ``This just in: Fake news packs a lot in title, uses
  similar, repetitive content in text body, more similar to satire than real
  news,'' in \emph{NECO 2017: The Second International Workshop on News and
  Public Opinion at ICWSM}, 2017.

\bibitem{faris2017partisanship}
R.~Faris, H.~Roberts, B.~Etling, N.~Bourassa, E.~Zuckerman, and Y.~Benkler,
  ``Partisanship, propaganda, and disinformation: Online media and the 2016 us
  presidential election,'' \emph{Berkman Klein Center Research Publication},
  vol.~6, 2017.

\bibitem{ribeiro2019auditing}
M.~H. Ribeiro, R.~Ottoni, R.~West, V.~A. Almeida, and W.~Meira, ``Auditing
  radicalization pathways on youtube,'' \emph{arXiv preprint arXiv:1908.08313},
  2019.

\bibitem{horne2019rating}
B.~D. Horne, D.~Nevo, J.~O'Donovan, J.-H. Cho, and S.~Adal{\i}, ``Rating
  reliability and bias in news articles: Does ai assistance help everyone?'' in
  \emph{Proceedings of the International AAAI Conference on Web and Social
  Media}, vol.~13, no.~01, 2019, pp. 247--256.

\bibitem{mena2019cleaning}
P.~Mena, ``Cleaning up social media: The effect of warning labels on likelihood
  of sharing false news on facebook,'' \emph{Policy \& Internet}, 2019.

\bibitem{salganik2006experimental}
M.~J. Salganik, P.~S. Dodds, and D.~J. Watts, ``Experimental study of
  inequality and unpredictability in an artificial cultural market,''
  \emph{science}, vol. 311, no. 5762, pp. 854--856, 2006.

\bibitem{vosoughi2018spread}
S.~Vosoughi, D.~Roy, and S.~Aral, ``The spread of true and false news online,''
  \emph{Science}, vol. 359, no. 6380, pp. 1146--1151, 2018.

\bibitem{horne2019different}
B.~D. Horne, J.~N{\o}rregaard, and S.~Adal{\i}, ``Different spirals of
  sameness: A study of content sharing in mainstream and alternative media,''
  in \emph{Proceedings of the International AAAI Conference on Web and Social
  Media}, vol.~13, no.~01, 2019, pp. 257--266.

\bibitem{benkler2018network}
Y.~Benkler, R.~Faris, and H.~Roberts, \emph{Network propaganda: Manipulation,
  disinformation, and radicalization in American politics}.\hskip 1em plus
  0.5em minus 0.4em\relax Oxford University Press, 2018.

\bibitem{norregaard2019nela}
J.~N{\o}rregaard, B.~D. Horne, and S.~Adal{\i}, ``Nela-gt-2018: A large
  multi-labelled news dataset for the study of misinformation in news
  articles,'' in \emph{Proceedings of the International AAAI Conference on Web
  and Social Media}, vol.~13, no.~01, 2019, pp. 630--638.

\bibitem{grover2016node2vec}
A.~Grover and J.~Leskovec, ``node2vec: Scalable feature learning for
  networks,'' in \emph{Proceedings of the 22nd KDD Conference}.\hskip 1em plus
  0.5em minus 0.4em\relax ACM, 2016, pp. 855--864.

\end{thebibliography}
\end{document}